\documentclass[useAMS,usenatbib]{mn2e}

\usepackage{graphicx}
\usepackage{txfonts}
\usepackage{natbib}
\usepackage{url}
\usepackage{color}
\usepackage{slashbox}
\usepackage{textcomp}
\catcode`\é=\active \defé{\'e}%
\title[Optical Turbulence Characterization at LAMOST Site]{Optical Turbulence Characterization at LAMOST Site: Observations and Models}
\author[L.-Y. Liu et al.]
{L.-Y. Liu$^{1,2}$\thanks{E-mail:liuly@nao.cas.cn}, C.Giordano$^{2}$, Y.-Q. Yao$^{1}$, J.Vernin$^{2}$, M. Chadid$^2$, H.-S. Wang$^1$,
\newauthor 
J. Yin$^1$, Y.-P. Wang$^1$\\
$^{1}$National Astronomical Observatories, Chinese Academy of Sciences, Beijing 100012, China\\
$^{2}$Universit\'e de Nice-Sophia Antipolis, Observatoire de la C\^ote d'Azur, CNRS-UMR7293, Laboratoire Lagrange, 06108 Nice Cedex 2, France}

\date{Accepted. Received; in original form}

\pagerange{\pageref{firstpage}--\pageref{lastpage}} \pubyear{2014}

\begin{document}

\maketitle

\label{firstpage}

\begin{abstract}
Atmospheric optical turbulence seriously limits the performance of high angular resolution instruments.
An 8-night campaign of measurements was carried out at the LAMOST site in 2011, to characterize the optical turbulence.
Two instruments were set up during the campaign:
a Differential Image Motion Monitor (DIMM) used to measure the total atmospheric seeing,
and a Single Star Scidar (SSS) to measure the vertical profiles of the turbulence
$C_n^2(h)$ and the horizontal wind velocity $V(h)$.
The optical turbulence parameters are also calculated with the Weather Research and Forecasting (WRF) model
coupled with the Trinquet-Vernin model, which describes optical
effects of atmospheric turbulence by using the local meteorological parameters.
This paper presents assessment of the optical parameters involved in high angular
resolution astronomy. Its includes seeing, isoplanatic angle, coherence time, coherence \'etendue,
vertical profiles of optical turbulence intensity $C_n^2(h)$ and horizontal wind speed $V(h)$.
The median seeing is respectively 1.01 arcsec,  1.17 arcsec and 1.07arcsec as measured with the
DIMM, the SSS and predicted with WRF model.
The history of seeing measurements at the LAMOST site are reviewed,
and the turbulence measurements in this campaign are compared with other astronomical observatories in the world.
\end{abstract}


\section{Introduction}
\label{sec:introduction}

The Large Sky Area Multi-Object Fiber Spectroscopic Telescope (LAMOST), is a special quasi-meridian
reflecting Schmidt telescope; its particular design allows both a large aperture
(effective aperture of 3.6 - 4.9 meter) and a wide field of view (FOV=5$^\circ$) (Richard 2008, Cui et al. 2012).
The main scientific goal of LAMOST focuses on the spectroscopic survey in the Galaxy and
extragalactic (Zhao et al. 2012). LAMOST located at Xinglong observatory of National Astronomical
Observatory of China(NAOC), which has been regarded as one of the important optical astronomical
observatories in China. The Xinglong Observatory is located in Hebei province, 170 kilometers northeast of
Beijing, where many telescopes are operating including the LAMOST, the 2.16 meter optical telescope and
several 1 meter telescopes.
A site testing campaign was performed in 2011 to obtain the atmospheric optical parameters
at Xinglong observatory, for the application of Adaptive Optics(AO), the active optics
and the high resolution spectrometer on the LAMOST and the 2.16 meter telescope(Zhang et al. 2004, Su et al. 2012).

The early work on optical turbulence characterization at Xinglong observatory
was carried out by Wu et al. (1996) and Song et al. (1998).
They carried out a 4 night campaign of multi-instrument measurements in December 1994.
The integrated seeing was measured with a DIMM, the surface layer turbulence was detected
with microthermal sensors and an acoustic radar.
Based on the photometry dataset in 1995-2001 of Beijing-Arizona-Taipei-Connecticut(BATC) telescope,
Liu et al. (2003) analyzed the Polaris images by measuring the Full Width at Half Maximum(FWHM)
to understand seasonal features of seeing condition.
Liu et al. (2010) performed DIMM seeing measurements on the top of LAMOST building,
in order to optimize the slit width of LAMOST spectrometer.
Yao et al.(2012) analyzed the seeing conditions with the BATC photometric data in 1995-2011,
and their results could be referred as the long-term (more than 15 years) seeing variation at Xinglong observatory.

In this paper we present the results of the site testing campaign at the LAMOST site in 2011.
In this campaign, two kinds of instruments were employed in order to obtain turbulence profiles
and integrated atmospheric parameters simultaneously. We have also calculated the atmospheric parameters
at the same time with a numerical method based on the WRF model coupled with the Trinquet-Vernin model (Giordano
2013, Trinquet 2007). In this paper we do not intend to show the reliability of WRF model to foresee a 3D map
of the optical turbulence due to too few days of observation. This goal was already demonstrated by
Giordano (2013).

In Section \ref{sec:instrument}, we briefly recall the optical turbulence parameters referring to
high angular resolution techniques, and describe the DIMM, SSS, and WRF model used in this campaign.
Section \ref{sec:results} presents the results obtained in the campaign.
The comparison and discussion of the results are given in Section \ref{sec:discussion},
in which we review the measurements at LAMOST site, and compare with other sites in the world.
Section \ref{sec:conclusion} is a summary.

\section{Instruments and Models}
\label{sec:instrument}


Atmospheric optical turbulence is the main factor of the degradation of image resolution. The major
characteristics of optical turbulence are described as profiles and integrated parameters. The two
main vertical profiles are the profile of the refractive index structure constant $C_n^2(h)$, and
the profile of the horizontal wind velocity $V(h)$.

In order to improve the techniques of high angular resolution (AO, spectroscopy, etc.),
one needs knowledge of the atmospheric integrated parameters, such as,
the Fried's radius $r_0$ or the seeing $\varepsilon_0$, the coherence time $\tau_0$, and the isoplanatic angle $\theta_0$.
The integrated parameters can be retrieved by the profiles above mentioned.

The Fried's radius is:

\begin{eqnarray}
r_0=0.185{\lambda}^{6/5}{\left(\int_{0}^{\infty}C_n^2\left(h\right)dh\right)}^{-3/5} ,
\end{eqnarray}

where $\lambda$ is the wavelength, $h$ is the altitude above the ground.

The seeing is one of the most important turbulence parameters to astronomy,
and usually the seeing value of observatory sites is around 1 arcsec in the visible band. As usual, here we
assume $\lambda=0.5$$\mu$m.

Seeing $\varepsilon_0$ (eq. \ref{equ:esp}), coherence time $\tau_0$  (eq. \ref{equ:tau}) and
isoplanatic angle $\theta_0$  (eq. \ref{equ:teta}),
can be derived from the vertical profiles of $C_n^2(h)$ and wind speed $V(h)$, as follows:

\begin{equation}
\label{equ:esp}
\varepsilon_0=0.98\lambda/r_0,
\end{equation}

\begin{eqnarray}
\label{equ:tau}
\tau_0=0.058{\lambda}^{6/5}{\left(\int_{0}^{\infty}{\left|V\left(h\right)\right|}^{5/3}C_n^2\left(h\right)dh\right)}^{-3/5},
\end{eqnarray}

\begin{eqnarray}
\label{equ:teta}
\theta_0=0.058{\lambda}^{6/5}{\left(\int_{0}^{\infty}{h}^{5/3}C_n^2\left(h\right)dh\right)}^{-3/5}.
\end{eqnarray}

The coherence \'etendue $G_0$ (Lloyd et al. 2004), is a comprehensive evaluation for adaptive optics with the combination
of Fried's radius $r_0$, isoplanatic angle $\theta_0$ and coherence time $\tau_0$, defined by the following formula:

\begin{equation}
 G_0=r_0^2 \tau_0 \theta_0^2.
\label{equ:G0}
\end{equation}

\subsection{DIMM}
\label{sec:DIMM}

The DIMM is a small transportable instrument,
with the differential technique to accurately measure the seeing conditions,
as described by Sarazin(1990) and Vernin(1995).
A small telescope is equipped with a mask made of two sub-apertures,
in order to obtain simultaneously two separated images of the same star on the focal plane.
Assuming a Kolmogorov model for the optical turbulence, the seeing can be calculated
from the variance of the differential motion between two star images.
Therefore, DIMM seeing monitor is almost insensitive to tracking errors, and the instrument
yields the integrated seeing from telescope pupil to the top of the atmosphere. Nowdays, the DIMM
technique is universally employed in site testing(Tokovinin, 2002; Aristidi, 2005).

According to the principle mentioned above, our DIMM instrument has the following configurations:

\begin{enumerate}
\item {a Meade LX200GPS telescope, with the entrance pupil of 20 cm, and the focal length of 200 cm;}
\item {a mask with two sub-aperture, with diameter of 5 cm, and a separation of 15 cm;}
\item {an optical wedge installed on one sub-aperture, to adjust an appropriate separation between the two star images at focus;}
\item {a Lumenera SKYnyx 2.0M CCD camera, attached to the telescope focal plane for fast sampling star images (\begin{math}\Delta t=0.5ms\end{math}),with a 640 $\times$ 480 format and the pixel size of 7.4 $\mu$m;}
\item {two computers, one calculating seeing values and closed-loop guiding, and the other remotely monitoring the status via internet.}
\end{enumerate}

\begin{figure}
\includegraphics[width=\linewidth]{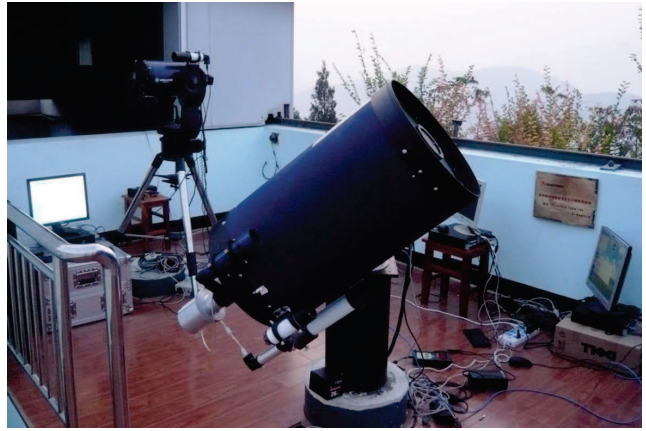}
\caption{The DIMM and SSS instruments are installed in the Public Dome, Xinglong observatory.}
\label{fig:dimm-sss}
\end{figure}

\subsection{Single Star Scidar}
\label{sec:SSS}

The Single Star Scidar is a new member in Scidar family (Vernin \& Roddier 1973; Fuchs,
Tallon \& Vernin 1998; Avila et al. 1998; Habib et al. 2006; Garc{\'{\i}}a-Lorenzo \& Fuensalida\ 2011;
Avil{\'e}s et al. 2012; Masciadri et al. 2012; Shepherd et al. 2014).
This instrument can deliver the distribution of $C_n^2$ with height, by using a small telescope.
The SSS technique analyzes the spatial-temporal auto- and cross-correlation
functions of less than 1 ms exposure images from a single star, which allow to retrieve the vertical profiles of
optical turbulence intensity $C_n^2(h)$ and horizontal wind speed $\textbf{v}(h)$ from the ground to the top of the atmosphere,
yielding the determination of seeing, isoplanatic angle and coherence time (See Section \ref{sec:instrument}).
The SSS technique in detail can be referred to Habib(2006) and Vernin(2009).
In order to obtain the contribution of surface layer turbulence,
the SSS also employs the so-called "Generalized Scidar" technique (Fuchs et al. 1998).
Although the SSS is a low-resolution optical turbulence profiler, it needs only 40 cm aperture telescope, showing a practical advantage
of being transportable for field site testing (Vernin et al. 2011, Giordano et al. 2012, Liu et al. 2012, Liu et al. 2013a).

In order to obtain optical turbulence profile at the LAMOST site,
we have constructed a new SSS with commercial components (Liu et al. 2013b).
The main specifications of our SSS are the following.

\begin{enumerate}
\item {a 40 cm Meade M16 tube, on an Astro-Physics 1200 equatorial mount, with the focal length of 400 cm;}
\item {a collimating lens after the focus of the telescope, with 10 mm focal length, to make the beam parallel;}
\item {a CCD camera, after the collimating lens for fast sampling the star scintillation pattern.
The fast readout Pixelfly CCD-200XS, 640$\times$480 format with pixel size of 9.9$\times$9.9 $\mu$m$^2$,
allows continuous acquisition with a high frequency rate without loss of any image.
The exposure time is usually taken as 1 ms, every 5.6 ms, the pixel scale is 0.51 arcsec/pixel};
\item {an auto-guiding system, made with a cube splitter;}
\item {two control computers, one capturing scintillation images from the CCD camera
and computing the spatial auto-correlation and cross-correlation images,
the other serving as real time tracking through a CCD guider.}
\end{enumerate}

\subsection{WRF Model}
\label{sec:WRF}

Some models have been developed in order to estimate the optical turbulence parameters from meteorological and topography
data (Coulman et al. 1988; Dewan et al. 1993; Masciadri et al. 1999a; Masciadri et al. 1999b; Trinquet \& Vernin 2007; Masciadri et al. 2013).
The Weather Research and Forecasting (WRF) model is a mesoscale non-hydrostatic numerical weather prediction system,
used for both operational forecasting and atmospheric research.
The WRF model allows to forecast the atmospheric parameters at each point of a given 3D grid,
such as the pressure P, the temperature T, the wind velocity components $\{$u,v,w$\}$ and the relative humidity (Skamarock et al. 2008).

The atmospheric optical parameters can be computed by the WRF model
coupled with the Trinquet-Vernin model (Trinquet H. \& Vernin J. 2006; Trinquet \& Vernin 2007; Giordano et al. 2014).
The Trinquet-Vernin model is a statistical model deduced from 160 meteorological balloons analysis,
and allows to compute the vertical profile of the $C_N^2$ from the vertical profiles of potential temperature and wind velocity.
A first study about the capability of WRF to predict the optical conditions is available in Giordano et al. (2013).

For the optical turbulence characterization during the campaign at the LAMOST site, the model was initialized as the following:

\begin{enumerate}
\item {horizontal coarse grid, composed of 100$\times$100 grid points,and spaced in $\Delta x=\Delta y$ = 27 km;}
\item {three consecutive nests with a ratio of 1/3, and therefore, the finest horizontal grid is $\Delta x=\Delta y$ = 1 km;}
\item {87 vertical levels from 0km to 20km, with a higher resolution within the low atmosphere $\Delta h_i$ = 50 m, and $\Delta h_{87}$ = 500 m;}
\item {the NCEP FNL (Final) Operational Global Analysis data (http://rda.ucar.edu/) was used to initialize the grid, and to set the boundary conditions.}
\end{enumerate}
The main parameters used for the simulation are:
\begin{enumerate}
\item {the microphysics scheme used is the Thomson scheme (mp\_physics = 8),}
\item {the Rapid Radiative Transfer Model (RRTM) scheme is used for the long wave radiation (ra\_lw\_physics=1),}
\item {the Goddard shortwave scheme is used for the shortwave radiation (ra\_sw\_physics=2),}
\item {the Mellor-Yamada-Janjic scheme is used for the planetary boundary layer (PBL) (bl\_pbl\_physics=2).}
\end{enumerate}

\section{SITE TESTING CAMPAIGN AND RESULTS}
\label{sec:results}

\begin{figure}
\includegraphics[width=\linewidth]{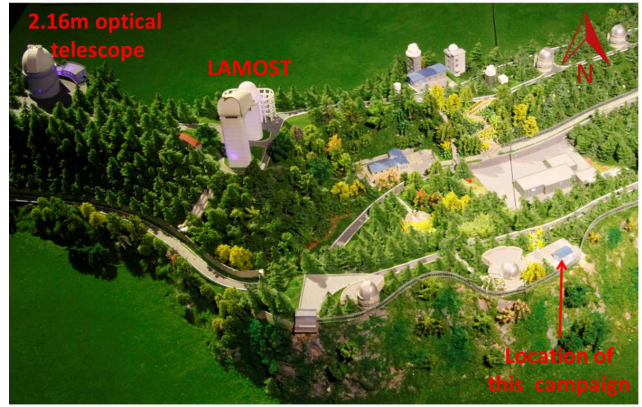}
\caption{The location of LAMOST on the Xinglong observatory (the photo was taken by Chen Yingwei).
The DIMM and SSS instruments are located 200 meters south-east of the LAMOST.
The 2.16 meter telescope is located at the north-west of the LAMOST.}
\label{fig:location}
\end{figure}

The site testing campaign was carried out over 8 nights from April 18 to September 26, 2011.
The DIMM and SSS were installed in the Public Dome of Xinglong observatory, 4m above ground level,
which is located about 200 meters south-east of the LAMOST building, as shown in Fig.\ref{fig:location}. The Public Dome can be completely open during observations.
The two instruments, separated by 3 meters, were operated with the same target simultaneously.
This configuration allows us to collect a substantial database and make cross-comparisons of the results.

\subsection{DIMM results}
\label{sec:DIMMresults}

\begin{figure}
\includegraphics[angle=270,width=\linewidth]{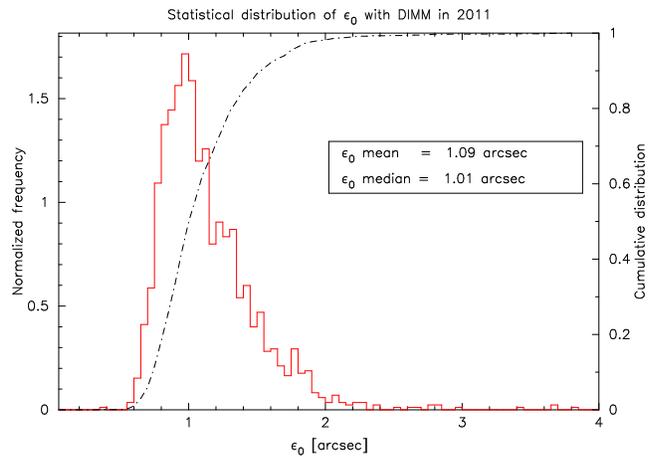}
\caption{The histogram (stairs) and cumulative histogram (dashed line)
of DIMM seeing values at the LAMOST site by the site testing campaign in 2011.}
\label{fig:dimm}
\end{figure}

\begin{figure*}
\includegraphics[angle=270,width=\linewidth]{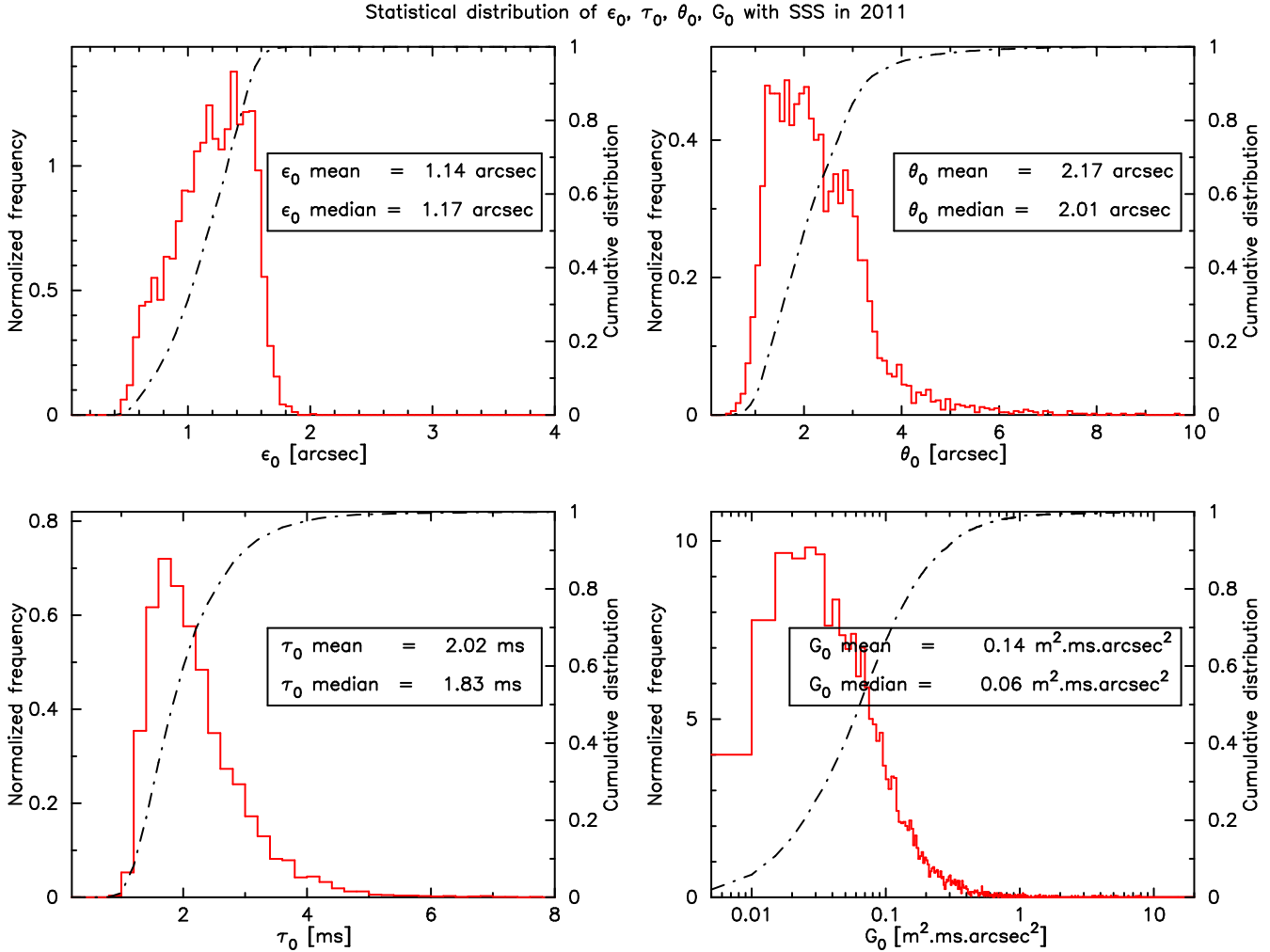}
\caption{Histogram (solid line) and cumulative distribution (dashed line) of atmospheric parameters
computed with the SSS measurements at the LAMOST site. The parameter distributions are
seeing $\varepsilon_0$ (top left),  the isoplanatic angle $\theta_0$ (top right),
the coherence time $\tau_0$ (bottom left), and coherence \'etendue $G_0$ (bottom right).}
\label{fig:sss_ao}
\end{figure*}

The DIMM seeing measurements were carried out from April 18 to 19 and from September 22 to 26.
A total of 1738 seeing measurements has been analyzed.
One DIMM seeing value was calculated
using a set of 100 images. The final seeing results have been corrected for the zenith angle.
The observing targets for the
DIMM seeing monitoring were $\alpha$ Leo (mV= 1.35 mag), $\alpha$ Boo (mV= -0.04 mag), $\alpha$ Lyr (mV= 0.03 mag),
$\alpha$ Aql (mV= 0.77 mag), $\beta$ Ori (mV= 0.12 mag), $\alpha$ Cas(2.24 mag) and $\alpha$ Cyg (mV$\approx$ 1.2 mag, variable star).

Fig.\ref{fig:dimm} shows the histogram and the cumulative distribution of the entire DIMM data set.
The mean seeing value is 1.09 arcsec, and the median value 1.01 arcsec.
There are 25$\%$ of the seeing values better than 0.9 arcsec, and 75$\%$ better than 1.2 arcsec.

\begin{figure*}
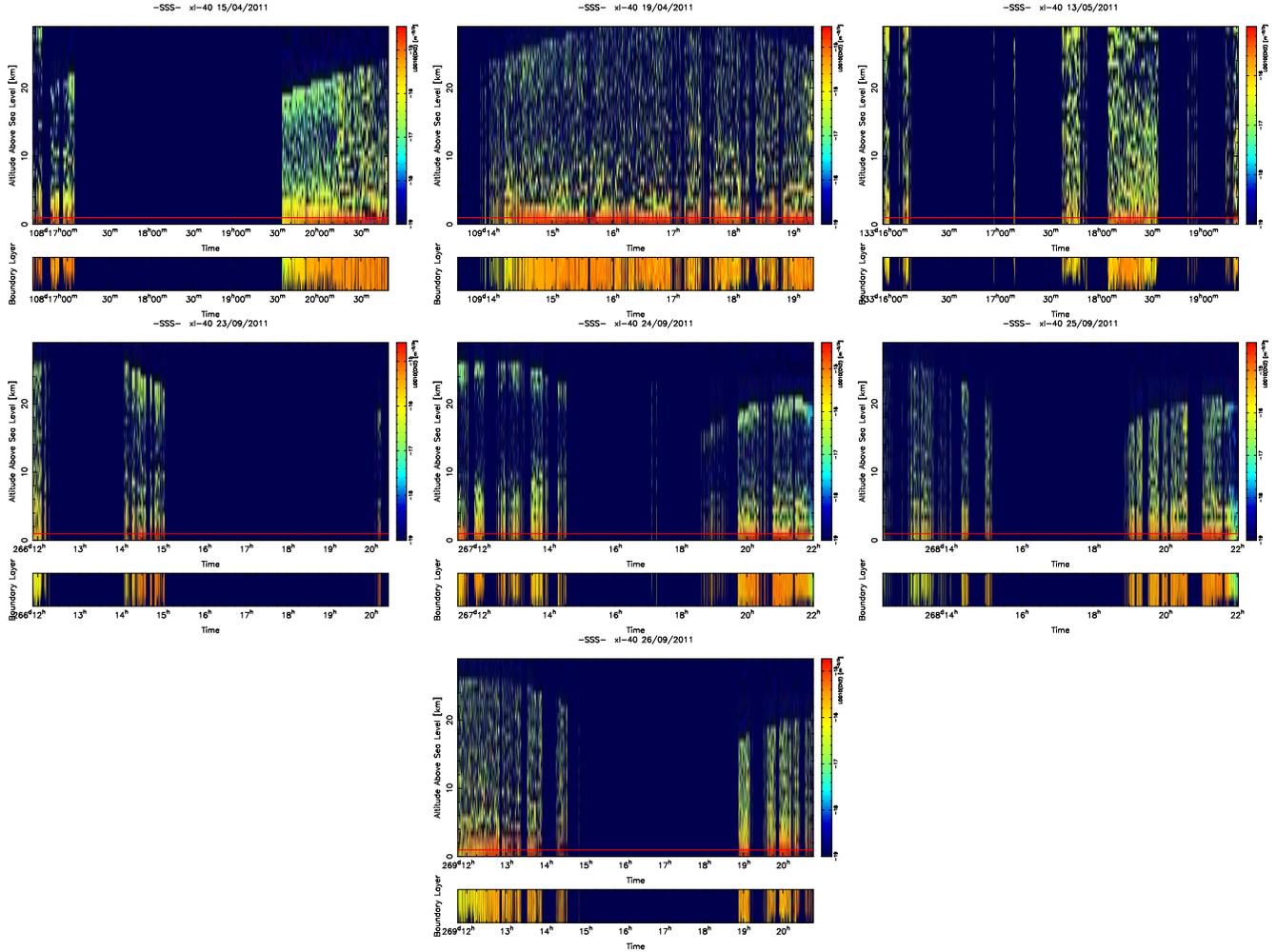

\begin{minipage}{0.33\linewidth}
  \centerline{\includegraphics[angle=270, width=0.98\linewidth]{f1.ps}}
\end{minipage}
\hfill
\begin{minipage}{.33\linewidth}
  \centerline{\includegraphics[angle=270, width=0.98\linewidth]{f2.ps}}
\end{minipage}
\hfill
\begin{minipage}{0.33\linewidth}
  \centerline{\includegraphics[angle=270, width=0.98\linewidth]{f3.ps}}
\end{minipage}
\vfill
\begin{minipage}{0.33\linewidth}
  \centerline{\includegraphics[angle=270, width=0.98\linewidth]{f4.ps}}
\end{minipage}
\hfill
\begin{minipage}{0.33\linewidth}
  \centerline{\includegraphics[angle=270, width=0.98\linewidth]{f5.ps}}
\end{minipage}
\hfill
\begin{minipage}{0.33\linewidth}
  \centerline{\includegraphics[angle=270, width=0.98\linewidth]{f6.ps}}
\end{minipage}
\vfill
\begin{minipage}{0.33\linewidth}
  \centerline{\includegraphics[angle=270, width=0.98\linewidth]{f7.ps}}
\end{minipage}
\caption{Temporal evolution of the $C_N^2$(h) during the 7 nights where SSS was observing. The
upper part of each plot has a one-km vertical resolution and the lower part refers to a kind of zoom
in the boundary layer (first km of the above plot). One can see that most of the time the first km (lower part of the
upper frame) is reddish, corresponding to $C_N^2\approx 10^{-15} m^{-2/3}$.
Within the free atmosphere, 1 km above ground level, the optical turbulence is faint and scattered from 1 to 30 km.
When $C_N^2$ profiles stops below 30 km, this means that the line of path of the star was far from zenith and,
 when straighten to the vertical, it was not possible to reach 30 km. See Habib(2006) and Vernin(2009) and references
 therein.}
\label{fig:sss_CN2}
\end{figure*}

\begin{figure}
\includegraphics[angle=270,width=\linewidth]{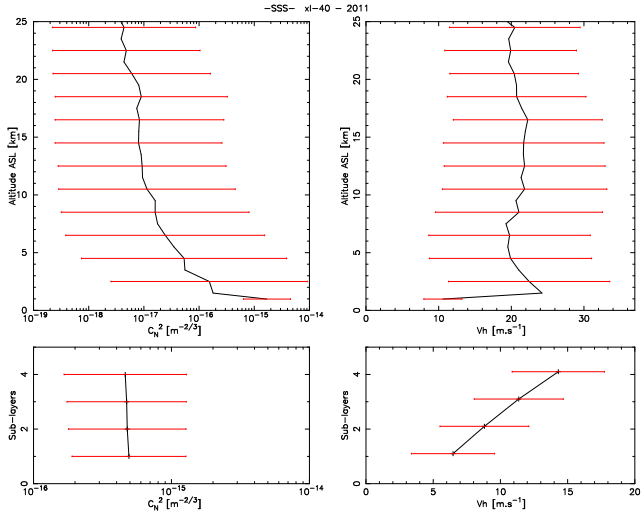}
\caption{The profiles of turbulence $C_n^2(h)$ and wind speed deduced by the SSS measurements at the LAMOST site in 2011.
The top-left is the median profile of turbulence $C_n^2(h)$ with $\pm\sigma_{log(CN2)}$ error bars,
and the top-right the median of the wind speed modulus profile $|V(h)|$ with $\pm\sigma$ error bars.
The average profiles of turbulence (bottom-left) and speed modulus (bottom-right) within the surface layer (4 sub-layers)
are also shown.
\label{fig:sssprofile}}
\end{figure}

\begin{figure}
\includegraphics[angle=270,width=\linewidth]{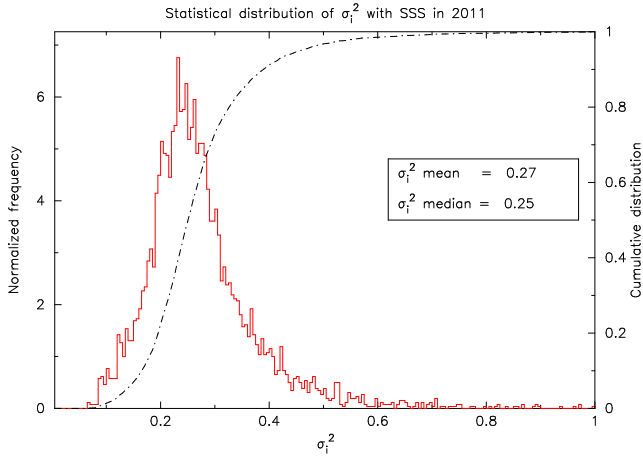}
\caption{Histogram and cumulative distribution (dashed line) of the scintillation index $\sigma^2_I$
measured by the SSS at the LAMOST site in 2011.}
\label{fig:sss_c0}
\end{figure}

\subsection{Single Star Scidar results}
\label{sec:SSSresults}

During the campaign in 2011,
a total of 6011 $C_n^2(h)$ and $|\textbf{v}(h)|$ profiles were obtained with the SSS from April 18 to 19, May 13, and from September 23 to 26.
The SSS spatiotemporal cross-correlation functions were obtained every 11.2 $sec$ corresponding to a set of 2000 images.
Due to measurement errors be caused by continuously changing sky background, especially in the night sky with moon,
the sky background changing with time was taken into account, and checked every hour during the SSS measurement.
The observing targets were $\alpha$ Aql (mV= 0.77 mag),
$\alpha$ Boo (mV= -0.04 mag), $\alpha$  Lyr (mV= 0.03 mag), $\beta$ Ori (mV= 0.12 mag) and $\alpha$ Cyg (mV$\approx$ 1.2 mag, variable star).

Fig.\ \ref{fig:sss_ao} presents the statistical distributions of the high angular resolution parameters.
The seeing values (top left) span over the range [0.4 and 2.0 arcsec],
with a half of the seeing better than 1.17 arcsec, and 75\% of the seeing better than 1.4 arcsec.
The isoplanatic angle (top right)
seems to follow a bimodal distribution with an average of 2.17 arcsec,
and a median of 2.01 arcsec, respectively. The coherence time (bottom left)
spans over a 1. to 5 ms range with
an average and median values of 2.02 and 1.83 ms, respectively.
For the coherence \'etendue (bottom right), the average and median values are 0.14, and 0.06 m$^2$.ms.arcsec$^2$.

Fig. \ref{fig:sss_CN2} presents, night by night, the temporal evolution of the optical turbulence over
the whole campaign, from ground level up to 30 km, along with detailed explanation given in the caption.

Fig.\ref{fig:sssprofile} shows the mean vertical profiles of turbulence and wind speed over the whole atmosphere.
Almost all the optical turbulence is concentrated within the boundary layer,
and the rest being scattered through the free atmosphere.
The wind speed, shown on the top right panel of Fig.\ref{fig:sssprofile},
increases from 10 $m.s^{-1}$ at ground level to reach 24 $m.s^{-1}$ at 2 $km$.
As discussed in Giordano et al. (2012), the SSS can retrieve four values of turbulence and wind speed within surface layer.
Due to its low vertical resolution, the SSS technique is not able to distinguish the altitudes of the four layers.
Typically, there is a wind gradient above a few hundred meters of Earth's surface.
Therefore, we arbitrarily sorted these layers
with increasing wind speed from bottom to top of the boundary layer, which seems reasonable and already
checked by Giordano et al. (2012).
The bottom panel of Fig.\ref{fig:sssprofile} shows the profiles of turbulence and wind speed modulus
within the first four sub-layers, corresponding to a surface layer (0-1 $km$) slab.
One can notice the steep wind speed gradient from 6 $m.s^{-1}$ to 14 $m.s^{-1}$ in the boundary layer.

The relative error of the correlation function is defined by Vernin et al. (1983):
\begin{eqnarray}
\frac{\Delta C}{C(0)}=\sqrt{\frac{1}{N_{speckles}}},
\end{eqnarray}
where $\Delta C$ is the estimation error, $N_{speckles}$ is the amount of independant atmospheric speckles analysed, and $C(0)$ is
the scintillation variance which is created by all of the layers. As is well known, the scintillation index for a given turbulent layer with altitude h as expressed by Roddier (1981):
\begin{eqnarray}
\Delta\sigma_i^2=\Delta C=19.2{\lambda}^{-7/6}{h}^{5/6}\Delta(C_N^2dh),
\end{eqnarray}
The SSS absolute error can be deduced with formula (6) and (7):
\begin{eqnarray}
\Delta(C_N^2dh)=0.052C(0){\frac{{\lambda}^{7/6}{h}^{-5/6}}{\sqrt{N_{speckles}}}}.
\end{eqnarray}

The scintillation index $\sigma_I^2$ (the normalized intensity variance) is related to the $C_n^2$
profile (Roddier 1982),

\begin{eqnarray}
\sigma_I^2=19.2{\lambda}^{-7/6}{\left(\int_{0}^{\infty}{h}^{5/6}C_n^2\left(h\right)dh\right)}.
\end{eqnarray}
where I is the normalized intensity fluctuation.
Here, the scintillation is measured on each point of the telescope pupil, which are $\approx0.7$ cm spaced,
much less than the first Fresnel zone which $\approx2 cm$ for a $1$ km altitude layer. One can assume that the pupil
averaging is almost negligible.
The autocorrelation of the scintillation $C_*(r)$
measured by the SSS on the pupil plane could deduce the scintillation index $\sigma^2_I=C_*(0)$.
Fig.\ref{fig:sss_c0} is the statistical distribution of the scintillation index at LAMOST site in 2011;
the mean $\sigma_I^2$ is 0.27, and the median $\sigma_I^2$ is 0.25.
That $\sigma_I^2<0.6$ for 99$\%$ of the scintillation index indicates the regime of weak fluctuations,
and allows to use the small perturbation theory (Roddier 1981).

%
%

\subsection{WRF model results}
\label{sec:WRFresults}

We used the WRF model to forecast the optical conditions at the same time as the DIMM and
SSS run. The WRF model was configured to compute one forecast every 10 minutes
in order to have a good temporal resolution.

\begin{figure*}
\includegraphics[angle=270,width=\linewidth]{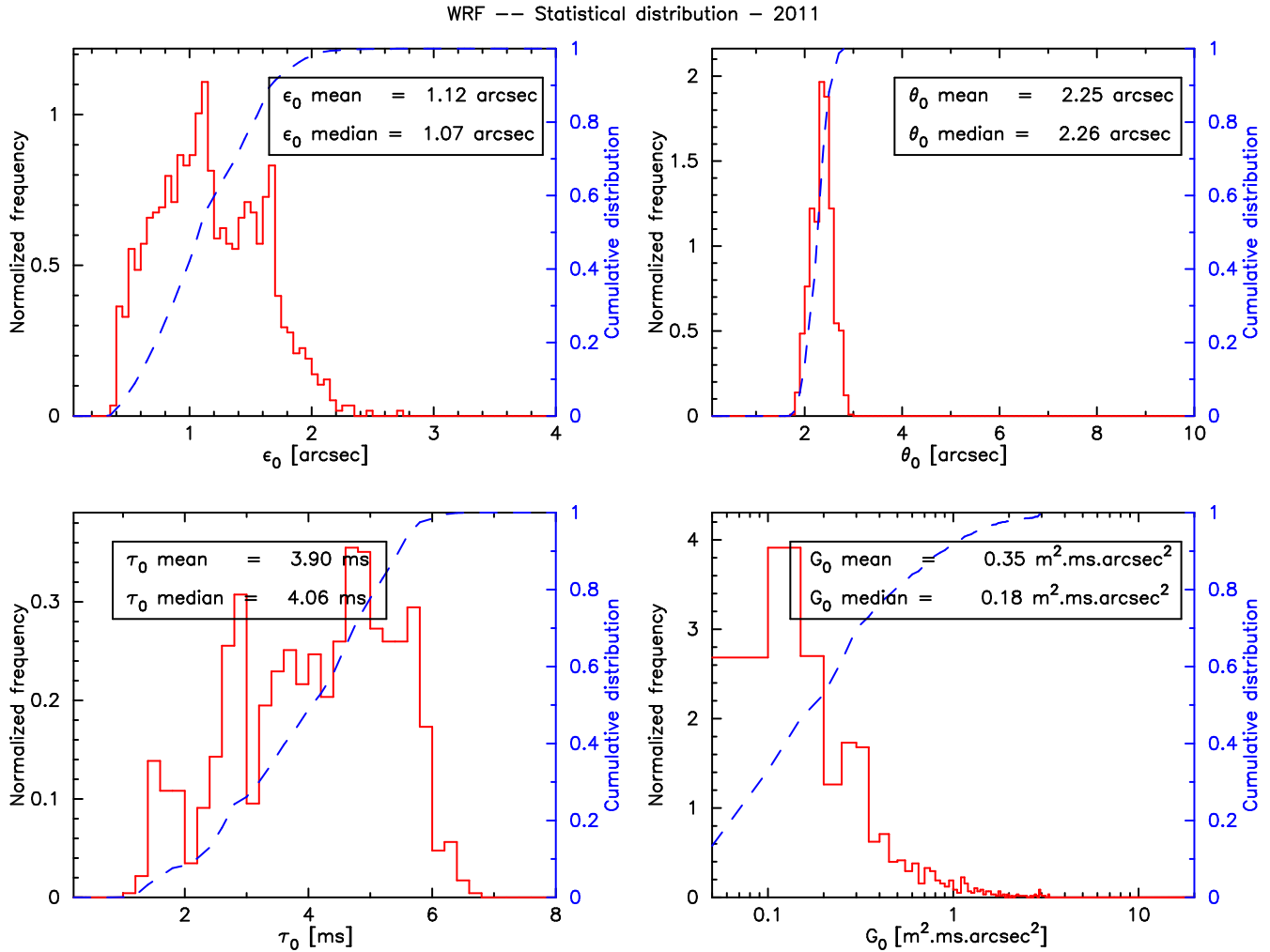}
\caption{Histogram (solid line) and cumulative distribution (dashed line) of atmospheric parameters forecasted with WRF model,
from April to September 2011 at the LAMOST site. }
\label{fig:wrf_ao}
\end{figure*}

Fig.\ref{fig:wrf_ao} shows the statistical distributions of the four integrated atmospheric optical parameters.
The seeing has a bimodal distribution with the first bump at 1.1 arcsec and the second at 1.6 arcsec.
The median and mean seeing values are 1.07 arcsec and 1.12 arcsec, respectively,
very close to the DIMM seeing values (1.01 arcsec and 1.09 arcsec), the standard deviations $\sigma$ is 0.28 arcsec.
Moreover, the cumulative distribution shows that 75\% of the seeing is better than 1.4 arcsec.
The range of the isoplanatic angle $\theta_0$ deduced from the WRF model is two times smaller than the SSS measurements,
but the median and mean values are close (2.26 arcsec and 2.25 arcsec with WRF model compared to 2.01 and 2.17 with the SSS,
and their $\sigma$ is 0.78 arcsec).
The distribution of the coherence time by WRF model is really different from the SSS measurements,
and the median and mean values are 4.06 ms and 3.90 ms, respectively, much greater than the SSS results.
The coherence \'etendue $G_0$ by WRF model has a log-normal shape similar as that by the SSS,
and the median and mean $G_0$ are 0.18 and 0.35 m$^2$.ms.arcsec$^2$, also close to the SSS values.

\begin{figure}
\includegraphics[angle=270,width=\linewidth]{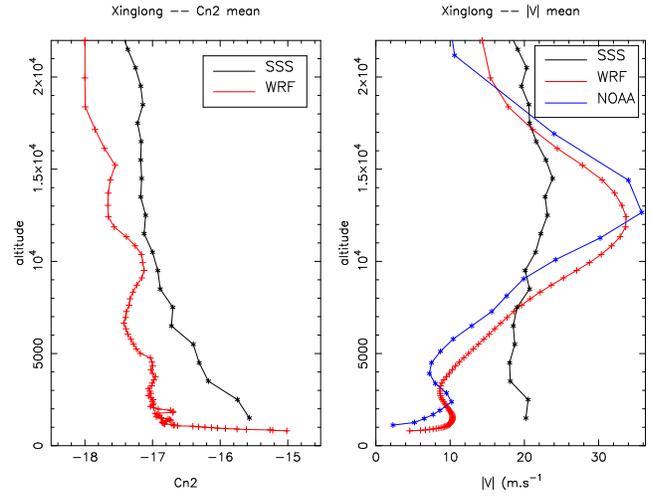}
\caption{The comparison of the average profiles of turbulence $C_n^2(h)$
and horizontal wind velocity $|V(h)|$ on 18 April 2011,
obtained by the SSS, by WRF model, and by the NOAA, the Air Resources Laboratory of National Oceanic and Atmospheric Administration.
The blue line by the NOAA shows the wind speed above Xinglong observatory at 21:00 UT on 18 April 2011.
\label{fig:wrf_profile}}
\end{figure}

Fig.\ref{fig:wrf_profile} presents the average vertical profiles of the $C_n^2(h)$ as measured by the SSS and by WRF model,
and the vertical profiles of the average horizontal wind velocity $\textbf{v}(h)$, by the NOAA, by the SSS, and by WRF model.
It seems that the profiles by WRF model do not meet those by the SSS measurements.
The $C_n^2(h)$ profile by the SSS is distinctly larger than that by WFR model;
the wind profile by the SSS shows higher velocity in lower atmosphere and lower around the tropopause,
and looks quite different from that by WFR model.
The vertical profile of wind velocity by WRF model and by the NOAA looks consistent, though there is about 1 km offset in altitude,
with a minimum at the ground level of 4 $m.s^{-1}$ and a maximum around 12.5 km of 35 $m.s^{-1}$.

\section{DISCUSSION}
\label{sec:discussion}

We have performed a campaign at the LAMOST site in 2011 to characterize the optical turbulence
with the instruments of DIMM and SSS, and by the WRF model.
The correlation analysis are performed with the results in order to compare the DIMM and SSS measurements,
as well as to check for the WRF model previsions, which have been already validated by Giordano (2013).

Comparing the seeing distributions as measured by the DIMM (Fig.\ref{fig:dimm}), by the SSS
(Fig.\ref{fig:sss_ao}) and WRF model (Fig.\ref{fig:wrf_ao}), one can notice that mean and median values are
coherent, but the shape of the distribution looks somewhat different. WRF seeing distribution shows a bimodal
behavior whereas DIMM peaks at low seeing values and SSS at large seeing values. If one assumes that free atmosphere
optical turbulence is steady, as seen in Fig.\ref{fig:sss_CN2}, and the boundary layer turbulence is more intermittent,
one should expect a bimodal distribution as already encountered by Aristidi et al. (2009) in Antarctica. But,
we have no explanation to the different DIMM and SSS seeing distribution.

\begin{figure*}
\includegraphics[angle=270,width=\linewidth]{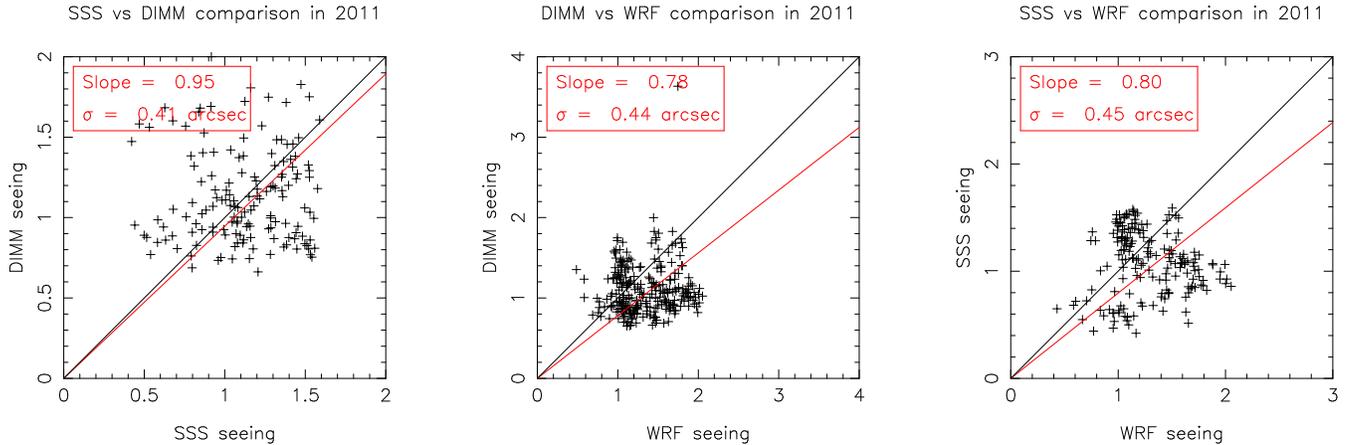}
\caption{The correlation maps of 10 minutes average seeing $\varepsilon_0$ for  SSS vs DIMM, DIMM vs WRF, and SSS vs WRF.
The black line means complete correlation, and the red line is the linear fitting. }
\label{fig:seeing_corr1}
\end{figure*}

\begin{figure*}
\includegraphics[angle=270,width=\linewidth]{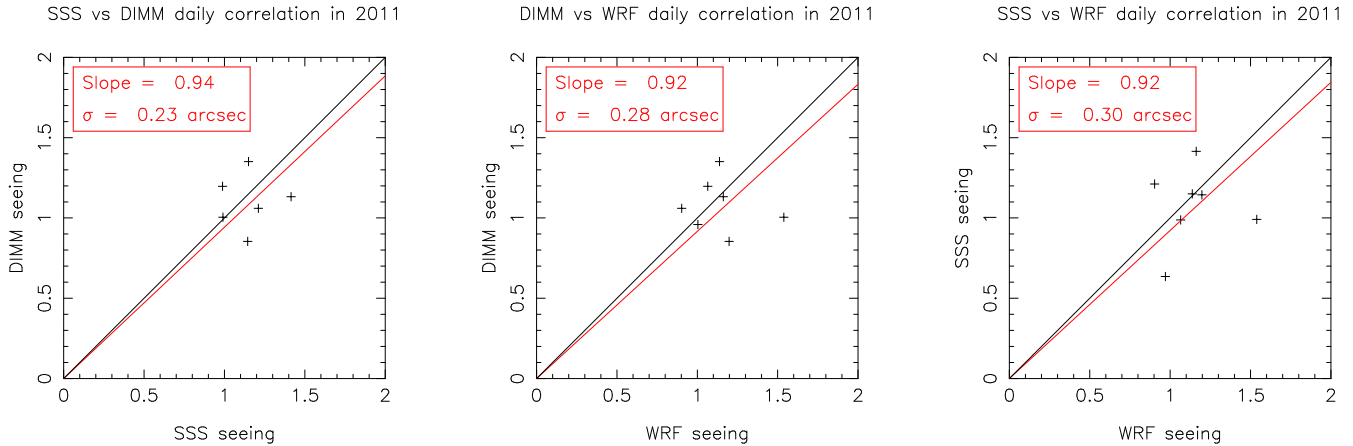}
\caption{Same as figure \ref{fig:seeing_corr1}, but for the seeing values averaged over one night.}
\label{fig:seeing_corr2}
\end{figure*}

\begin{figure*}
\includegraphics[angle=270,width=\linewidth]{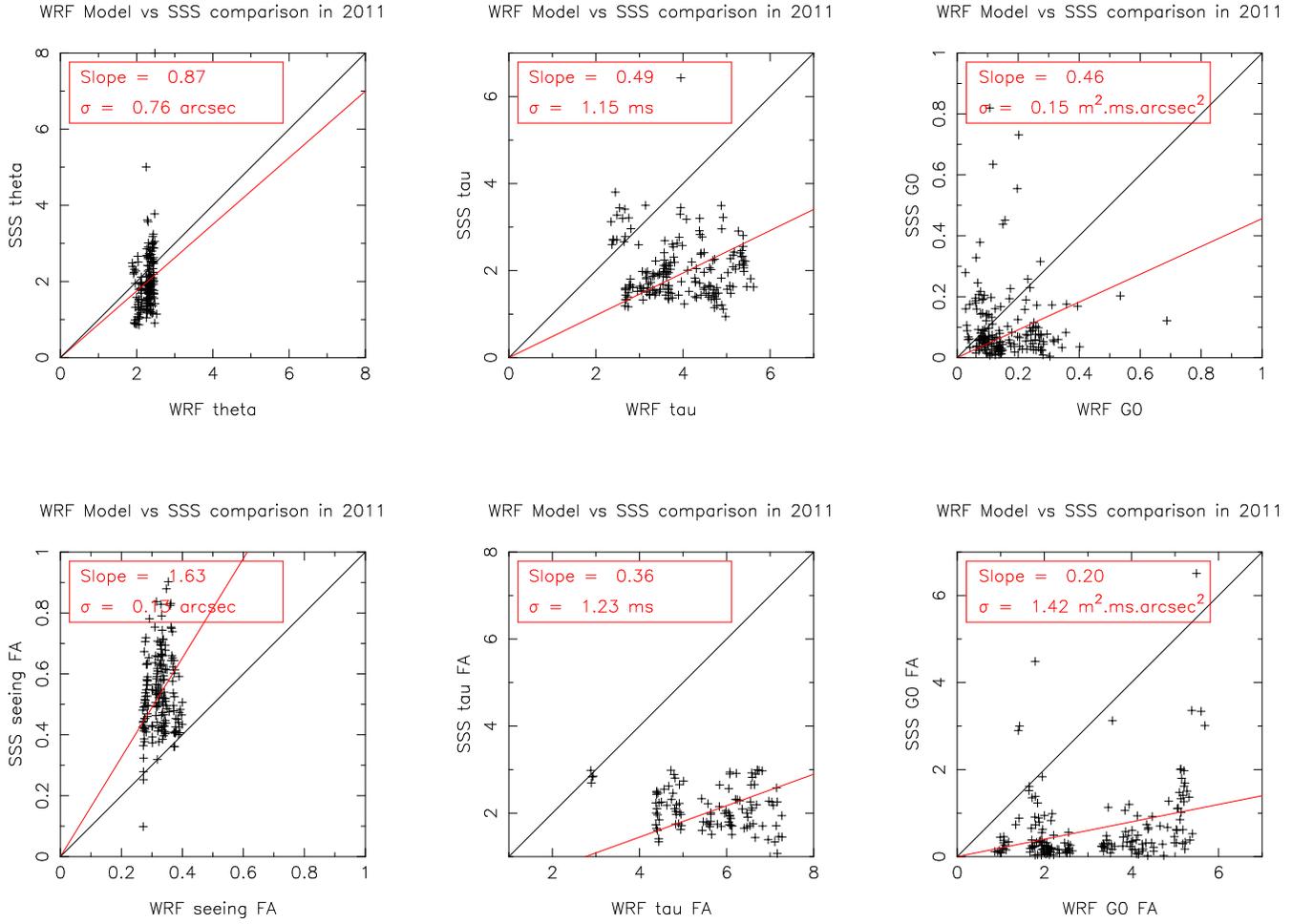}
\caption{The correlation maps of 10 minutes average optical turbulence parameters measured with the SSS and forecasted with WRF model. The upper diagrams are the isoplanatic angle $\theta_0$, the coherence time $\tau_0$ and the coherence \'etendue $G_0$, and the below are the respectively parameters of free atmosphere (FA): the seeing $\varepsilon_{FA}$, $\tau_{FA}$ and $G_{FA}$.
The black lines mean complete correlation, and the red lines indicate the linear fitting.}
\label{fig:ao_corr}
\end{figure*}

Fig.\ref{fig:seeing_corr1} shows the comparisons of the seeing averaged every 10 minutes by the three methods.
Statistically, DIMM and SSS seeings
agree within a standard deviation of $\sigma$=0.4 arcsec. WRF seems to predict larger
seeing values than SSS and DIMM, with a regression slope of 0.8 and 0.78.
Even if the three regression slopes are close to one, meaning that there is no systematic bias,
one can notice a large scatter.
Indeed, the standard deviations of the linear fitting show $\sigma$=0.41 arcsec between the DIMM
and SSS seeing, $\sigma$=0.44
arcsec for the DIMM and WRF seeing, and $\sigma$=0.45 arcsec for the SSS and WRF seeing.
The correlation between the SSS and DIMM seeing seems to be a little better.
For site characterization, a 10 minute resolution seems too high, and a one day seems a more
appropriate sampling.
Fig.\ref{fig:seeing_corr2} further compares the seeing values averaged over each night.
One can see that not only the regression slope is closer to one, but the dispersion $\sigma$ is almost two
times better. If one assumes that SSS and DIMM are well characterizing the seeing, WRF seems
a good tool for seeing characterization, as already mentioned by Giordano et al. (2013) and discussed
in section \ref{sec:WRF}.

The comparisons of the other turbulence parameters measured by the SSS and calculated by WRF model
are shown in Fig.\ref{fig:ao_corr}, for
the isoplanatic angle $\theta_0$, the coherence time $\tau_0$ and the coherence \'etendue $G_0$,
and also the parameters for free atmosphere (FA): $\varepsilon_{FA}$, $\tau_{FA}$ and $G_{FA}$.
We find that the agreement is not good; though for the seeing
$\varepsilon_{FA}$ the deviation is small ($\sigma$=0.13 arcsec),
for the $\tau_0$ and $G_0$, it is not the case.
The isoplanatic angle is dominated by high altitude (FA) layers which are predicted within a small
dynamic range by WRF. WRF optical turbulence relies on Trinquet-Vernin parametrization which is based upon
vertical gradient of the potential temperature and wind speed, but the vertical spacing is coarse. This means
that FA dependent parameters such as $\varepsilon_{FA}$ and $\theta_0$ are not subject to large variations.

Tab.\ref{tab:stat} compares the optical turbulence parameters at the LAMOST site with
those at ORM, Mauna Kea, Armazones, and the Antarctica sites as Dome C and South Pole,
(Vernin et al. 2011, Sch\"{o}ck et al. 2009, Giordano et al. 2012, Marks et al. 1999).
We find that the isoplanatic angle $\theta_0$ at the LAMOST site is similar to the other sites,
but the seeing and coherence time at the LAMOST site are poorer.
The poor coherence \'etendue $G_0$, at the LAMOST site, as well as South Pole, may be not good for adaptive optics.
This can be due to the high optical turbulence distribution in the boundary layer, associated to a steep wind shear, from 6 up to 14 $m.s^{-1}$ in the first four ground layers (See Fig.\ref{fig:sssprofile}, the bottom right panel).

\begin{table*}
\begin{tabular}{lccc|ccc|cc}
\hline\hline
& & & \multicolumn{2}{c |}{Site} \\
\cline{2-7}
\centering Parameters & ORM  & Mauna Kea & Armazones & Dome C & South Pole & Xing Long\\
\hline
Total seeing $\varepsilon_0$ (arcsec)          & 0.80 & 0.75 & 0.64 & 1.00 & 1.60 & 1.17 (SSS) \\
Isoplanatic angle $\theta_0$ (arcsec)          & 1.93 & 2.69 & 2.04 & 6.90 & 3.23 & 2.01 (SSS)\\
Coherence time $\tau_0$ (ms)                   & 5.58 & 5.10 & 4.60 & 3.40 & 1.58 & 1.83 (SSS)\\
Coherence \'etendue $G_0$ (m$^2$ ms arcsec$^2$)& 0.38 & 0.62 & 0.49 & 1.80 & 0.07 & 0.06 (SSS)\\
Reference to data & Vernin (2011) & Sch\"{o}ck (2009) & Sch\"{o}ck (2009)  & Giordano (2012) & Marks (1999) & This article \\
\hline\hline
\end{tabular}
\caption{The global median values of optical turbulence parameters.}
\label{tab:stat}
\end{table*}

\begin{table*}
\small
\begin{tabular}{lcccccccl}
\hline\hline\noalign{\smallskip}
Period &  Method &  Night &   Elevation(m) &  Spring(arcsec) &   Summer(arcsec)  &  Autumn(arcsec)  & Winter(arcsec)   &  Reference     \\
\hline\noalign{\smallskip}
Dec.1994        & DIMM &   4    &     15      &         &          &         &  1.4    &  Song (1998)\\
1995-2001       & FWMH &  All   &     6       & 3.5     & 2.9      &3.7      &  3.9    &  Liu (2003) \\
Oct.2003        & DIMM &   3    &     15      &         &          &         &  1.3  ¡¡&  Zenno (2004)\\
Mar.-Apr.2007   & DIMM &   12   &     28      & 1.1     &          &         &         &  Liu (2010)\\
Apr.-Sep.2011   & DIMM &   7    &     4       & 0.91    &          &1.09     &         &  This article\\
                & SSS  &   7    &     4       & 1.32    &          &1.10     &         &  \\
                & WRF model &8  &     -       & 1.12    &          &1.04     &         & \\
\noalign{\smallskip}\hline\hline
\end{tabular}
\caption[]{ The seasonal comparison of the median seeing measured with multi-methods at different locations of Xinglong observatory.}
\label{tab:history}
\end{table*}

In Tab.\ref{tab:history}, we recall all of the site testing campaigns at Xinglong observatory, and we collect the seeing results every measurements in different place,
in order to provide a reference to the application of Adaptive Optics(AO), the active optics and the high resolution spectrometer.
Song et al. (1998) measured the seeing in 1994, using a DIMM
inside the 2.16 m telescope dome, and the results could be affected by the dome seeing.
Liu et al. (2003) and Yao et al.(2012) analyzed the FWMH of Polaris images obtained by a Schmidt telescope
for a long period of time (1995- 2011). Although the results includes many influence factors,
such as the dome seeing, image quality, and tracking accuracy of the telescope, however, the data collection for over a decade is useful to understand more about the seasonal seeing conditions.
Liu et al. (2010) performed DIMM seeing measurements on the top of LAMOST building in 2007,
and obtained a better seeing with higher elevation than that at the 2.16m telescope.
In this work, we selected an open flat roof for the measurements, and for the first time,
we obtained whole night optical turbulence profiles;
this work should provide more reliable results to characterize the site condition at Xinglong observatory.

\section{CONCLUSION}
\label{sec:conclusion}

A site testing campaign has been carried out in 2011 at the LAMOST site
in order to characterize the optical atmospheric turbulence.
A DIMM and a Single Star Scidar have been set up during the campaign, and
the WRF model is applied to calculate the site condition and compare with the observation results.
For the first time, we are able to measure the turbulence profiles at Xinglong observatory.

The results of the optical turbulence parameters - seeing, isoplanatic angle, coherence time,
and coherence \'etendue, are presented.
The integrated seeing measured by the DIMM gives a median value of 1.01 arcsec.
There are 6011 individual profiles of optical turbulence $C_n^2(h)$ and horizontal velocity $|\textbf{v}(h)|$
recorded by the Single Star Scidar, and the median SSS seeing is measured to be 1.17 arcsec,
the median isoplanatic angle 2.01 arcsec, and the median coherence time 1.83 ms.
The global parameter coherence \'etendue $G_0$,
which is usually employed to evaluate the performance of high angular resolution,
gives a mean value of 0.14 and a median of 0.06 m$^2$.ms.arcsec$^2$.

The optical turbulence parameters are calculated over the same period of the campaign
with the WRF model, coupled with the Trinquet-Vernin model,
based on the global weather data by National Center for Atmospheric Research (NCAR).
The WRF model calculation gives a median seeing of 1.07 arcsec,
a median isoplanatic angle of 2.26 arcsec, and a median coherence time of 4.06 ms.
The coherence \'etendue $G_0$ by the WRF model gives a mean value of 0.35
and a median of 0.18 m$^2$.ms.arcsec$^2$. We find the WRF model can provides more accurate
seeing forecast on one night average than ten minute average.

If WRF previsions seems to depart from SSS/DIMM measurement, this could be due to the
fact that the $1x1$km terrain model might be to coarse when compared to the steep slant of about
200m slope situated along a SW-NE direction ridge, at the SE border of XingLong observatory.

The optical turbulence parameters obtained in the campaign are compared with
other sites in the world, and the seeing measurements in history at Xinglong Station are reviewed.
The Coherence time $\tau_0$ (1.58 ms) and Coherence \'etendue $G_0$ (0.07 m$^2$ ms arcsec$^2$) at LAMOST
site is poor, leading to the conclusion that no exceptional conditions should be expected
for adaptive optics.
The $C_n^2(h)$ measurements presented in this work are evaluated over only a short period.
More DIMM and SSS measurements are needed in the future for a comprehensive optical turbulence evaluation
of LAMOST site.

\section*{ACKNOWLEDGMENTS}

This work is supported by the National Natural Science Foundation of China
(NSFC, Grant Nos. 10903014, 11373043 and 11303055), and the Young Researcher Grant of National Astronomical
Observatories, Chinese Academy of Sciences. We thank the Public Dome of Xinglong observatory for
their support during the SSS and DIMM measurements. We also thank the National Center for Atmospheric Research for
its access to weather archive, and the WRF community for access to these software and for his help during
its utilization.

\bibliography{xinglong}   
\label{lastpage}
\end{document}